\documentclass[12pt]{article}
\usepackage[utf8]{inputenc}
\usepackage{epsf}
\usepackage{url}
\usepackage{graphics,color}
\usepackage{amsmath}
\usepackage{amssymb}
\usepackage{yfonts}
\begin{document}
\begin{center}\Large{\bf A Golden Age of General Relativity?\\\large
    Some remarks on the history of general relativity.}\\ 
\normalsize Hubert Goenner\\ Institute for Theoretical Physics\\ University
  G\"ottingen\\ Friedrich-Hund-Platz 1\\37077  G\"ottingen\end{center} 
%Status 12. 7. 2016
\section{Introduction}
In 1986, Jean Eisenstaedt has published a well received paper: ``The
low water mark of general relativity: 1925-1955'' on activities concerning
Einstein's general relativity \cite{Eisen1986}, \cite{Eisen2006}. The
first date coincides with the year of the break through to quantum
mechanics by Werner Heisenberg, accompanied by Max Born and Pascual Jordan, with Schr\"odinger following them in 1926 via a different
route. It also signifies the year from which on Einstein's interest
shifted from general relativity to his unsuccessful unified field
theory. The second date possibly is not intended to point to Einstein's
death year but to coincide with the Jubilee-Conference in Bern, in
1955. For D. Kennefick who called Eisenstaedt's period ``an
interregnum in research in general relativity'', the advent of quantum
mechanics and the paucity of experiments in gravitation were
sufficient reasons for an explanation (\cite{Kenne2007},
p. 105). It was only consequent that Clifford C. Will, at the end of
the 1980s, wrote of a ``Renaissance of General Relativity'': ``by the
late 1950s general relativity had become a sterile, formalistic
subject, cut off from the mainstream of physics. [..] Yet by 1970,
general relativity had become one of the most active and exciting
branches of physics.'' (\cite{Will1989}, p. 7). Other authors set the
date for the demise of general relativity even a decade earlier: ``In
the 1940s the subject of general relativity was virtually dead - or at best
dormant.\footnote{It is unclear whether Bergmann's book of 1942 on general relativity was on the mind of the authors \cite{Berg1942}, or rather not.} Peter  
[Bergmann] set up the first active research group in GR in the
USA. Singlehandedly he resurrected the field. Several years later
other schools of GR developed around J. Wheeler, H. Bondi, L. Infeld
and P. Jordan, but it is clear that Peter was the first to understand
the importance of resurrecting GR from its dormancy and placing it as
an important part of fundamental physics and then actively pursuing
this goal'' \cite{SaGiMe2003}. Also, Robert Dicke was considered to
have played a role: ``He soon became a leading figure in what is known
as `renaissance of general relativity' ''\cite{Kragh2016}.\\  

Another claim differs from these statements assuming a period of
reduced activity in the field of general relativity and its
applications: Kip Thorne wrote of ``the golden age of general
relativity'' as ``the period roughly from 1960 to 1975 during which
the study of general relativity, which had previously been regarded as
something of a curiosity, entered the mainstream of theoretical
physics'' (\cite{KipTh2003}, p. 74). Ten years earlier, he had given
an even more precise period, i.e., from 1963 to 1974
\cite{KipTh1994}. \\ 

It is obvious that, after World War II, research on general relativity
and other relativistic theories of gravitation broadened enormously
both in the number of workers and the spread of universities which housed ``relativists''; cf. e.g., table 1 in (\cite{BluLaRe2015}, p. 614). My
aim here is to show that a label like ``renaissance of general
relativity'' rests on weak empirical ground while ``the golden age of general relativity'' is an exaggerated description of a period of rapidly growing activity. For historiography, both labels should be used with caution.\\ 

In order to assess such historiographic catchwords, a more precise
definition of ``Golden Age'' must be given. With regard to this
concept we might ask whether it is to describe: \\   
 - A period of great, publicly visible activity in general relativity?\\ 
 - A period which brought great {\em theoretical} advances for general
relativity?\\  
 - A period which brought great advances in {\em experimentation} and
{\em observation} applicable to test general relativity?\\
Similar questions may be raised with regard to the ``period of
marginalization within the field of physics''\cite{BluLaRe2016}, also
coarsened to signify the ``dark ages'' of general relativity
\cite{Schu2012}, both alternative interpretations for the ``low water
mark'':\\ 
 - Is the assumed low-level status of general relativity within the physics community from 1925 to the 1950s well grounded in the conceptual development of the theory?\\
 - If such a period existed, to what degree was it due to reasons {\em external} to science proper?\\ 
 - Do we have to distinguish different developments in different countries?\\ 
For the establishment of both claims beyond personal interests, we
may ask whether:\\ 
 - the authors quoted above have used any {\em quantitative}
indicator?\\
   - a naive comparison of the periods 1925-1955 and 1955-1975 is
meaningful at all?\\  

In the following we will discuss and try to answer some of these
questions. We shall distinguish between {\em communications} (public
awareness of the field, teaching, conferences) and {\em achievements}
(conceptual progress, publication of research, institutions).\\ 

\section{Manpower, funds, and activity-indicators} 
Before we can evaluate the development within the field of general
relativity, we must look at physics as embedded in the larger area of
society. As we know, physics was notably influenced by the impact
of World War II, and by the ensuing ``Cold War''.  It is common knowledge that World War II, e.g., due to the development of radar, rockets, or of the atomic bomb, by many was branded as ``the physicists war'' \cite{Kaiser2015}. Keywords for some effects correlated with
politics are the ``increase in manpower in physics [including general
relativity]'' and the new ``private and military funding of physics
[including general relativity]''. 
\subsection{Manpower and financing}
Since September 1956, the US Air-force
intensely supported research in gravitation through the ``General
Physics Laboratory of the Aeronautical Research Laboratories (ARL) at
Wright-Patterson Air Force Base, Dayton, Ohio'' (\cite{Witten1998},
p. 375). This financial source was brought to an end in 1969 by the Mansfield Amendment \cite{Goldberg1992}. To my knowledge, the
supported projects were not classified. In Europe, since 1958, the
North Atlantic Treaty Organization (NATO) also supported science by
both individual grants, by its conference series and the subsequent book publications. Although, in the past, only a trickle has flown into grants for gravity research, the program continues until this day \cite{NATO2015}.  

At the same time, roughly, funds from industrial companies went into
research in gravitation. Exemplary is the ``Research Institute for
Advanced Studies''  (RIAS), Baltimore, founded in 1955 by George
Bunker of the Glenn L. Martin Company. The RIAS changed its name and
turned its mission away from basic research in 1973. Another example
is the manufacturer {\em Texas Instruments}. The university of Texas
at Dallas began in 1961 as a research arm of Texas Instruments, and
with it soon the ``Southwest Center for Advanced Studies'' (SCAS). In
1969 the founders bequeathed SCAS to the state of Texas. As is known,
SCAS became a nucleus for gravitational physics, brought into flower
mostly by European scientists holding a permanent job there
(I. Robinson and W. Rindler from England/Vienna),  and by some from
Pascual Jordan's Hamburg Relativity-Seminar with shorter engagements
like those of J. Ehlers (1964-1965) or C. Bichteler  (1966-1968, in
the mathematics department); the last two moved on to the University
of Texas in Austin.\footnote{ Both, Ehlers (from 1966 to 1971) and
  Bichteler (1969 to retirement)}\\    

Another point to be made is the difference in {\em manpower} of
physicists [relativists] working during the ``low water''-  and the
``renaissance'' -periods. Since the launch of Sputnik, in 1957, a
shortage of physicists in the USA was claimed, and an advantage on the
manpower front assumed for the Soviet Union (\cite{Kaiser2002},
p. 148). Countermeasures in the educational system were taken. In the
report of the Physics Survey Committee of the National Academy of
Sciences of the USA we read: ``[..] between 1964 and 1970, the number
of PhD physicists increased by 60 percent, the number of PhD
astronomers by 62 percent, and the number of PhD's in astrophysics and relativity\footnote{This should not be misunderstood as refering to
PhD's produced; physicists {\em holding} PhD's are meant.} by 300
percent (from 65 to 257 individuals)'' (\cite{NatAcad1973},
p. 840-841). Also, around 1970 the number of physicists in the  
subfield ``astrophysics and relativity'' amounted to only 1.5 \% of
the total of physicists with a PhD in the USA. This number is
mentioned here in order to put into perspective the claim that,
between the 1950s and 1970, general relativity was ``one of the most
active [..] branches of physics.''  Around 1970, there were three times
more astronomers  n o t  working in relativity. Although these figures
refer to the USA, we may assume that a similar if less pronounced
situation existed in (NATO-) Europe. We note that a quantitative
comparison of the periods pre- and post-1955 could clearly only be made per head of scientists working on relativistic gravitation. A corresponding study has not yet been done.\\ 
Naturally, manpower and funding are closely correlated.  This is shown
clearly  by Fig. 1, relating Federal R \& D expenditures and
bachelors' student production in the physical sciences, mathematics
and engineering in the United States (\cite{APSN2003},
p. 8):   \begin{center} \includegraphics{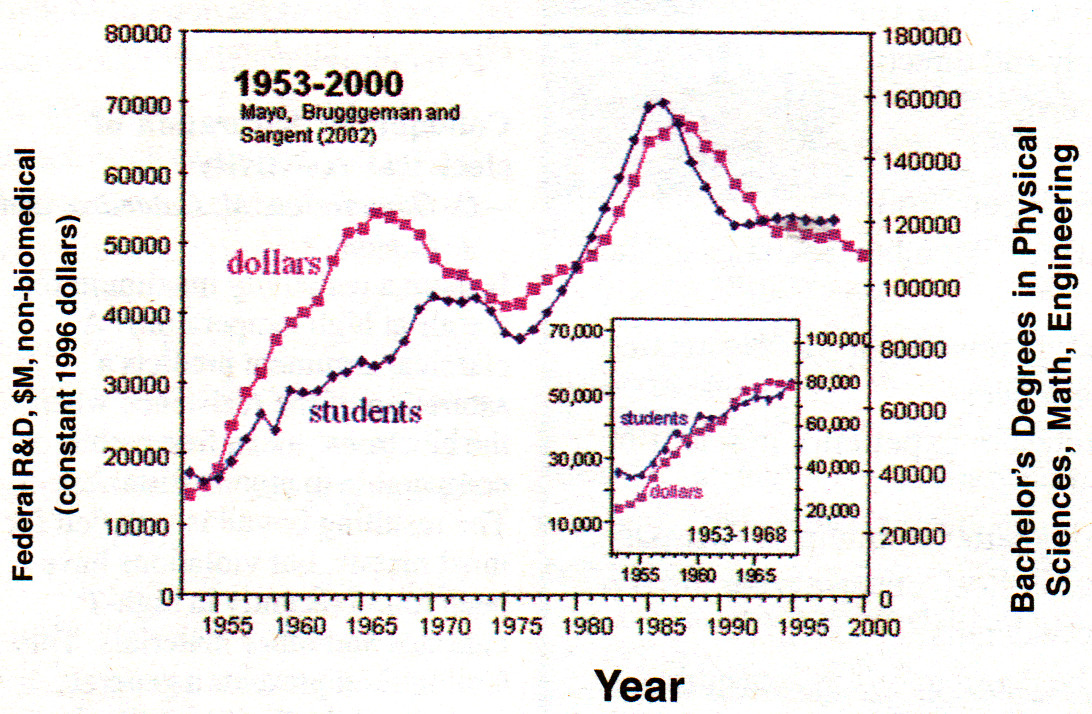}\\ Figure 1. \end{center} 
 
The number of  bachelor students almost doubled between 1957 and 1973
with the expenditures increasing by a factor 2 1/2 ! This remains
unparalleled in the period 1925 to 1955; it is explained by reasons
{\em external} to the field of gravitation. An indication that, at
least in West-Germany, research in gravitation developed {\em
  smoothly} during the three decades from 1960 to 1990, is the number
of PhD’s produced there in the field of general relativity. It amounts to
equally 13-14 for each decade. Thus, to a great extent what was named
the ``Golden age of relativity'' in the United States, may have been nothing but
a feature of a general trend in physics after the
``Sputnik''-shock. \\                                                                                  
                                                                                                                                 
\subsection{Activity indicators}
As activity-indicators, we could take the number of national and
international conferences organized, the number of groups/single
persons working in relativity {\em relative to some reference group},
the number of books and papers published, the number of journals
regularly printing papers on general relativity, or centers with a
graduate program in general relativity. We also note that
the necessary funds for international conferences including the travel
costs had not been available before the 1950s.\footnote{Of course, all such
indicators do have their weaknesses \cite{LarIns2010}.}\\

\subsubsection{Publications}
Not one of the indicators suggested above was checked by those
claiming a ``Golden Age'' of General Relativity. Looking at the
bibliographies by Combridge at King's College \cite{Comb1965} 
\begin{center}
\includegraphics{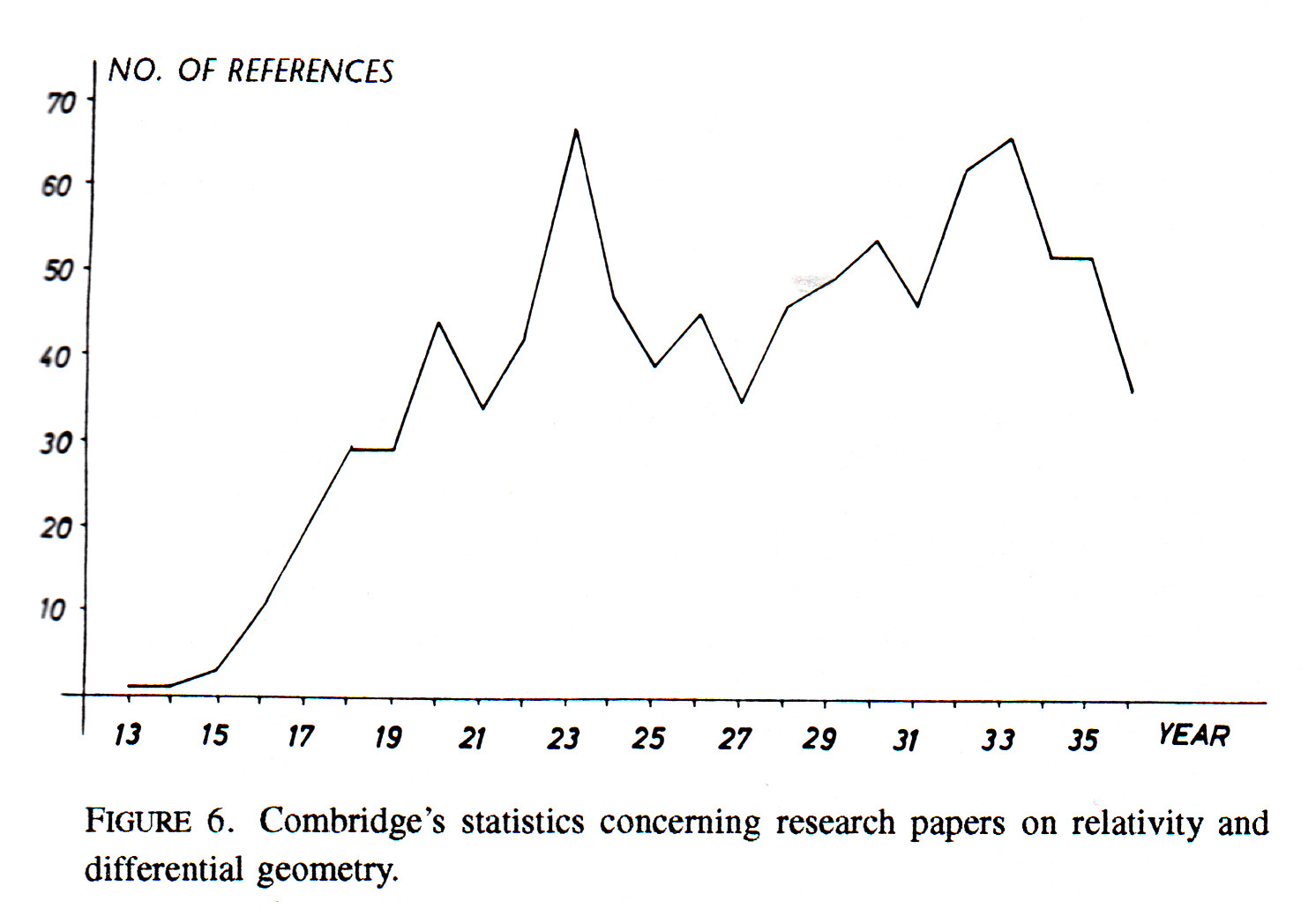}
\end{center}
and in Synge's book on general relativity \cite{Synge1960}, the average
yearly output of papers amounted to $\simeq 14/year$, and from 1925 to
1955 to roughly $12/year$, whereas between 1955 and 1965 the rate of
published papers rose to ca. $15/year$. Papers on general relativity
published in the prestigious French journal Comptes Rendus grew from a yearly total of 10 in 1957 to a maximum of 23 in 1959 and stayed at
about this yearly rate until 1964. Although not representative, these
inspections do not point to remarkable variations in publications
listed between 1915 and 1960, in particular not between 1925 and 1935,
or in the first half of the 1960s. \\ 
J. Eisenstaedt , in his first article of 1986, gave the yearly number
of papers in relativity for the five years 1932-1936 to be around
sixty. This figure goes well with Combridge’s statistics.\footnote{He
  did not say whether papers on special relativity are in- or
  excluded. His data were taken from Physics Abstracts and
  Physikalische Berichte.} He also gives the number of 30 articles on
general relativity published in 1955 (from Physics Abstracts).\\  
This is in stark contrast to the stable growth rate in the number of
physics publications {\em in all fields of physics} between 1920 and 1960, an
exponential growth, doubling the number of papers in approximately 15
years \cite{SolPri1963}. Of course, exponential growth need not be
present in special fields of physics as shown by a study on weak
interactions \cite{SuWhiBa1977}. In another study three periods of growth in
scientific publications (all fields) are given: compared with less than 
1\% growth before, to 2 to 3\% up to the period between the two world wars and 
8 to 9\% to 2012 \cite{BorMu2014}. Some {\em representative} empirical
material for the time after 1945, may be obtained from the core
collection of the ``web of science''.\footnote{We have searched for
  the topics ``general relativity'' and ``Einstein's theory''. From
  the latter list all references not related to general relativity have
  been removed.} From 1945 up to today, there was a continuous
growth of publications in general relativity with the steepest
increase between 1960 to 1970, 1995 to 2000, and 2010 to
2015.\footnote{The 5-fold increase in the decade after 1990 possibly
  is due to a changed data base used by the web of science (more
  journals included in the database?).}  
\begin{center} \includegraphics{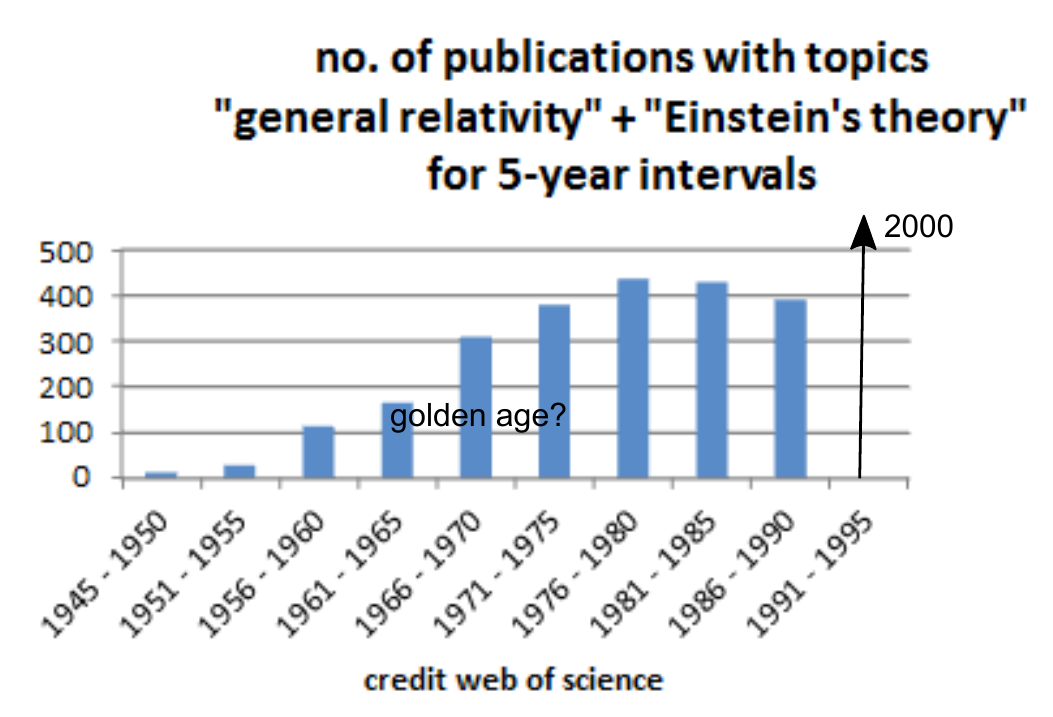}\\ Figure 3. \end{center} 
If the past ``Golden Age'' is identified with the steepest growth of 
publications on general relativity, then the period of 1960 to 1970 is
the correct one. 
. 
 \begin{center} \includegraphics{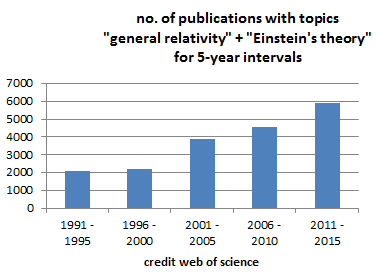}\\ Figure 4. \end{center}

\subsubsection{International Conferences}
As to conferences {\em before} world war II, there existed very few such
events. Among them were the Volta Conferences in Italy by the
Royal Academy of Science in Rome. The first such conference in 1927 was the 
Como Conference, held at Lake Como in 1927 about the uncertainty
principle by Niels Bohr and Werner Heisenberg. A well known older
series of Conferences were the Solvay Conferences in Brussels. We note
those conferences relevant for relativity and gravitation:\\ 
- 1911 ``La th\'eorie du rayonnement et les quanta; The theory of
radiation and quanta'' (Einstein present);\\  
- 1958 	La structure et l'\'evolution de l'univers (The structure and                      evolution of the universe);\\ 
- 1964 The Structure and Evolution of Galaxies;\\  
- 1973 Astrophysics and Gravitation.\\
Before the 1940s, international conferences could be financed only by scientific academies and wealthy entrepreneurs (like Ernest Solvay).
Due to the increase in state funds for science and education, from the
mid of the 1950s, national and international conferences on
gravitation and cosmology, partially as continuing series, began to
sprout and became a regular feature of activity in the field of
gravitation. An (incomplete) list of the best known such conferences
is given in Appendix 1. If taken as an indicator for activity, the
organization of conferences on gravitation fulfills the claim that the
period starting around 1960 to 1975 brought both the weaving of a net
of ``relativists'' and the awareness for the field in the physics community.\\
However, due to the lack of funds and the lower amount of working
power in physics, such an increase in conferences could not have
happened during 1925 up to the decade after World War II. In this
context, the ``low water mark of general relativity'' seems to be a
projection from the present into the past. 

 \section{Advances in research about general relativity} 
Since the 1920s, during Einstein's lifetime and thereafter, a number
of important results were obtained. For
briefness, we present a choice of them - after 1920 - chronologically in four lists. The first collects results from {\em the time before world war
  II}: \begin{quote} 
{\bf 1923/24  Exact solutions describing an expanding universe} with
space sections of constant curvature (Friedman);\\
{\bf 1925 -1933 The universe as an exploding primordial atom}
(Lema$\hat{i}$tre) with a beginning later derisively named ``the big
bang'';\\ {\bf 1929} Hubble-law.\\
{\bf 1936/1937 Gravitational lensing}; theory by
  A. Einstein. Application to galaxy clusters by F. Zwicky.\\
{\bf 1938 Equations of motions for point particles in linear
   approximation.} Unlike in Maxwell's theory, the equations of motion
 in the gravitational field cannot be postulated independently of the
 field equations (Einstein-Infeld-Hoffmann).\\ 
{\bf 1939 Gravitational collapse (Oppenheimer, Snyder, Volkoff)
  Tolman- Oppenheimer-Volkoff-limit (TOV).} This triggered a
development leading eventually to the concepts of ``white dwarfs'',
``neutron stars'' both as stellar remnants, and of ``black holes''
(treated as solutions of the vacuum field equations).\end{quote} 
\noindent In the {\bf next list}, results obtained {\em after world
  war II} until the mid 1950s are assembled: 
\begin{quote}{\bf 1946/48  The theory of Cosmic Background
    Radiation. (Dicke, Gamov, Alpher \& Herman)} It is one of the
  pillars of present cosmology.\footnote{Its first observation occured
    in 1964 (Penzias \& Wilson); that this radiation has a black-body
    spectrum corresponding to a temperature of  approximately 2.725 K
    was convincingly measured much later.} \\ 
{\bf 1949  Gödel-cosmos (K. Gödel)} This is a locally rotating exact
solution of Einstein's equations with dust matter. Universal space
sections do not exist, but closed timelike worldlines.\\ 
{\bf 1954/58  Petrov-Pirani classification. (Petrov 1954, Pirani
  1958)} An application is the peeling theorem describing the
asymptotic behavior of the Weyl tensor in a lightlike direction.\\  
{\bf 1956/58  Event horizon (Rindler, Finkelstein)} W. Rindler
formulated the concept within cosmology (event- and particle
horizons). Finkelstein applied it to a collapsing star and showed that
an event horizon develops.\\ 
Gödel, Petrov, and Penrose are mathematicians, Dicke, Gamov, Alpher,
Herman, and Finkelstein physicists.\end{quote}
\noindent The {\bf third list} contains progress {\em during the
  1960s} achieved in general relativity. It falls into the categories:
mathematical physics without empirical backing, new mathematical
methods, and the application of general relativity to astrophysics. 

\begin{quote} {\bf 1960/62  Spinors: Newman-Penrose formalism.} A technique for
formulating general relativity with spinors in place of tensors
introduced by R. Penrose and E. T. Newmann. The formalism is  helpful
for the characterization of outgoing gravitational radiation in
asymptotically Minkowskian space time. \\ 
{\bf 1962   Exact solutions as an initial-value problem (Arnowitt-Deser-Misner, ADM)} Einstein's field equations are decomposed into hyperbolic time-evolution equations and elliptic constraint equations. The formalism is also important for numerical calculations.\\
{\bf 1963   Kerr-metric (Roy Kerr)} (Exact solution describing the field outside of a  star rotating about a fixed axis. As no interior solution (with matter) is known, the Kerr- metric is interpreted as a black hole with 2 parameters: mass and angular momentum.\\
{\bf 1965  Kerr-metric with electric charge (Newman)} (A rotating black hole with 3 parameters: mass, angular momentum, electric charge. It is a solution of the Einstein-Maxwell (vacuum) equations.) \\
{\bf 1964  Popularization of the name ``Black Hole'' (Wheeler)}
{\bf 1966-1970 Singularity theorems (Penrose, Hawking)}\end{quote}

\noindent The {\bf fourth list} collects some of the important developments of the 1970s. Note that most of them concern purely theoretical statements and conjectures with no empirical basis.

\begin{quote}{\bf 1960 - 2000 Uniqueness theorems for stationary Black Holes (Israel, Carter, Mazur)} ``A black hole has no hair'' (J. Bekenstein). \\
{\bf 1969 Cosmic-censorship-hypothesis (Penrose)} Singularities are always hidden behind event horizons; otherwise, the theory would become unpredictable.\\
{\bf 1970/71  Post-Newtonian approximation (Nordveth, Will)} It is the modern form of the EIH-method and is used to describe observable effects and to discern alternative gravitational theories.\\
{\bf 1970/73 Gravitation as Poincar\'e gauge theory (F. Hehl).} 
{\bf 1972 Black-Hole-Thermodynamics (J.   Bekenstein)}, A hypothesis about a relation between the area of the event horizon of a black hole and thermodynamical entropy is introduced. The main theorem of thermodynamics are postulated, analogously.\\
{\bf 1974 Hawking-radiation of a Black Hole} By a hypothetical quantum mechanical effect, a black hole, which by definition cannot send a signal from the inside through the event horizon, now can radiate its energy away.\\ 
{\bf 1975 Begin of numerical relativity} (L. Smarr; axial symmetry, 2 dimensions) 
\end{quote}
Note that the important exact solutions for a rigidly rotating disk formed by dust \cite{NeuMein1993}, \cite{NeuMein1995}, or its generalization to differentially rotating disks \cite{AnsMein2000} were found only during the 1990s.\\
From the preceding enumeration, it looks as if conceptual and
methodical progress in the understanding of general relativity and of
its consequences has continued {\em uninterrupted} and {\ em unweakened - with
somewhat smaller growth rates from 1920s to the 1960s than from the
1960s until 1975. In my view, these small differences do not warrant the
conclusion that a low water mark for the period 1925 to 1955 existed. 
\section{A new field: Relativistic astrophysics}
What usually is not in the focus of the proponents  claiming a ``renaissance of general relativity'', is the birth of the new field of
``Relativistic Astrophysics'' since the 1950s, particularly from the
beginning of the 1960s on \cite{Schueck1989}. The need for an explanation of observations on quasars (1963), pulsars (1967), neutron stars (1971), binary pulsars (1974), with the indirect evidence of
gravitational waves, led to the introduction of general relativity
into astronomy. Moreover, with the theory and subsequent observation
of the cosmic background radiation, cosmology no longer remained an
academic subfield  of general relativity but became part of
relativistic astrophysics.\footnote{As a matter of fact, in 1992 the
  astrophysicist David Schramm used the label ``A Golden Age of
  Cosmology'' refering to the much later period beginning with the
  1990s \cite{Schra1992}.} It is questionable, whether the many papers
on ``relativistic astrophysics'' and on ``cosmology'' can be simply
subsumed under ``general relativity''. In any case, they contributed
substantially to the increasing output of publications in relativistic
gravitation during the 1960s and 1970s - without constituting a renaissance:
``relativistic astrophysics'' just did not exist before the 1950s.  
\section{Universities with sizeable production of PhDs in
  general relativity after world war II} 
We concentrate on the years after world war II until the 1950s. Three
places stand out: Princeton with John A. Wheeler (1911-2008), Syracuse
with Peter G. Bergmann (1915-2002) \cite{Newman2005} and Hamburg with Pascual Jordan (1902-1980) and Otto Heckmann (1901-1983). While R. Feynman and A. Wightman obtained their PhD degrees with J. Wheeler already in the 1940s (unrelated to gravitation), 5 PhDs followed in the 1950s, 3 in the 1960s, and 4 during the 1970s.\footnote{These are lower limits; there might have been further doctorates in general relativity with Wheeler.} Seven fall into Thorne's ``Golden-Age''-period. As Clifford Will and Daniel Kennefick were PhD-students of Kip Thorne, they might have been inclined to share the enthusiasm of their adviser. Interestingly, those of Wheeler’s students with PhDs related to gravitation before 1960 like A. Komar, D. Brill, H. Everett, or Ch. Misner did not speak of a ``Golden Age''. Likewise, from PhD-students of Peter Bergmann like Ralph Schiller, Rainer K. Sachs and Joshua N. Goldberg, I have seen no statements about ``renaissance'' or ``Golden Age''. The same is true with Jordan's (Heckmann's) students J. Ehlers, W. Kundt, M. Trümper, E. Schücking. The spawning of doctor's degrees from the three theoretical physicists Bergmann, Jordan and Wheeler gradually brought into existence a net of well known relativists, and through them led to growing activity on relativistic theories of gravitation. Since the mid 1950s, one of such became the regular
``Stevens Relativity Meeting'' which received its name from the
Stevens Institute of Technology in Hoboken, N.J., where R. Schiller and
J. Anderson taught. An institutionalising of the field was still far
away.\footnote{In the American Physical Society, a topical group 
  ``gravitation'' was established not before 1995.} Of course, in the
1960s, in many further places research on general relativity was taken
up like in London (Bondi, Hoyle), Warsaw (Infeld), Austin (Schild), Philadelphia (P. Havas) etc. In view of these developments, it is understandable when Wheeler’s student Kip Thorne felt like living in golden times. In his obituary for Wheeler, D. Overbye  reformulated his role as expected: ``He rejuvenated general relativity; he made it an experimental subject and took it away from mathematicians [..]''\footnote{One of those  mathematicians might have been V. Hlavat\'y of Indiana University.} \cite{Over2008}.\\ 
Who could have advised doctoral students in general relativity {\em before} the second world war?  Einstein never cared for PhD-students and never had one, in Berlin and in Princeton. De Sitter, Lorentz (PhD student J. Droste), Max von Laue, Thirring, and mathematicians Hilbert, Klein and Weyl? For all of them, general relativity was not central to their work, and was not the reason for their fame. Nevertheless, M. v. Laue had several doctoral students in special and general relativity, from e.g., Ernst  Lamla (PhD 1912 on special relativistic hydrodynamics) \cite{Lamla1912} to Max Kohler (PhD 1933 on general relativistic optics and cosmology) \cite{Kohler1933}. This has to be considered if periods before and after 1950 are compared and a ``low-water-mark in dissertations'' could be asserted. 

\section{Are the arguments for the  ``Low water mark'' and ``Golden Age'' of General Relativity convincing?}
From 1915 until the 1990s, general relativity amounted to ``Little
Science'' in the terminology of de Solla Price\footnote{``Little
  Science'' is done by single researchers and small groups with modest
  monetary resources. ``Big Science'' involves a large number of
  collaborators and large funding.} \cite{SolPri1963}. Compared to the
overall physics community, workers on relativistic gravitation never
surpassed 1\% to 2\% of all physicists. Without a definition of what
{\em mainstream} is to mean, dates for the entrance of general
relativity into the mainstream are unconvincing. Does it mean that
more universities have hired physicists working in relativistic
gravitation? In Germany this {\em never} did happen. Does it mean that
more courses on general relativity were held? This would be
unspecific: with the increase in the workforce in general relativity
and astrophysics, more people could and did give regular classes in
general relativity and/or cosmology.\footnote{The  fact that
  J. Wheeler held the first such course in Princeton only in 1952
  despite Einstein’s presence there since 1933/1934, seems due to
  Einstein's reluctance toward holding courses.} Eisenstaedt’s remark
that general relativity was at its peak in the 1920s
(\cite{Eisen1989}, p. 280) is true only if public awareness of the
theory is considered ({\em media hype}) \cite{Goenner1992}, not in
terms of research. As shown in section 3, in the years thereafter,
many conceptual advances equal to or even outclassing the previous
body of results and alleviating the understanding of the theory were
made. According to B. Schutz \cite{Schu2012}, while such advances
occured only within the mathematical ``skeleton'' of the theory,
heuristic concepts, all taken from relativistic astrophysics like
``black hole'', ``gravitational waves'' or ``gravitational lenses''
were still missing until the 1970s.\footnote{The concept ``graviton''
  was present, though.} For B. Schutz they are necessary for communication with
``non-relativists''. In this context, he asserts that general 
relativity became a ``complete theory'' only in the 1970. In view of 
the many discussions among philosophers of science as to when a
scientific theory may be called ``complete''\footnote{Is quantum
  mechanics complete? Many would negate this due to open questions
  around measurement. Cf. also \cite{Scheibe1973}.}, the introduction of such a concept is not helpful. 
A similar point of view may be found in \cite{BluLaRe2016} where the
uncertain ``epistemic status'' of general relativity theory (during the
low-watermark-period) is emphasized. The discussions about
the epistemic status of a quantum state again show that this concept
leads away from physics into the philosophy of science. \\
In the end, it is the lack of  experiments, the pull of quantum
theory, and the approximative Newtonian approach in the decades
between the first and the second world war that are mostly offered to
explain the low-watermark-period. \\ 
At present, in addition to the three effects from general relativity,
well known since the 1920s, half a dozen new effects have been
observed: Time delay effect, cosmic background radiation, frame
dragging, de Sitter geodetic precession, gravitational lensing effect,
gravitational waves. Only the first two were observed in 1964, i.e.,
within the period claimed as ``the golden age''.\footnote{The
  experiment by Dicke, Roll, and Krotkov  about the equivalence of
  inertial and gravitational mass also was published in 1964.} 
 In spite of being known since the 1920s and 1930s, all the other
 predictions had to wait for 60 years until the 1980s to 1990s and
 longer (until 2015) to be measured.  If there has been a
 low-watermark-period, then {\em for experimentation and observation
 related to general relativity}, and it has nothing to do with the asserted
 period 1920 to 1955.\\

From the first bets on the direct discovery of gravitational waves in
the mid 1960s (\cite{DiWheel1967}, p. 53) to their first observation,
fifty years have passed - notwithstanding ``renaissance'' and ``Golden
Age''. Resistance to pouring big funds into the possible observation
of gravitational waves persisted until the 1980s (\cite{Schu2012},
p. 261). Today, the only subdiscipline of general relativity deserving
the name ``Big Science'', is research related to gravitational
waves.\footnote{The use of Earth-Satellites on which experiments are
  flown for the test of general relativity could also be called ``big
  science''.} With the new field of ``relativistic astrophysics'' the
application of general relativity to objects astronomers previously
had looked at, mainly boosted research on relativistic theories of
gravitation. Quasi-stellar sources, pulsars, and neutron stars were
not known in the period interpreted as constituting the ``low water
mark''. Black Holes (stellar and galactic), gravitational lensing and
gravitational waves were bearing fruit {\em in astrophysics}. Probably, the emergence of relativistic astrophysics has helped generating the feeling of a ``Golden Age'' for general relativity .\\

It seems to me also that the route quantum mechanics has taken since
1925/1926, was unconsciously used as a model for how general relativity {\em should have developed}. In view of the scarce number of gravitational effects beyond those following from Newtonian theory and for which general relativity is needed, this theory could not claim to be of central importance to the physics community - fundamental issues related to space and time notwithstanding. 
\section{Conclusion}
Writing the history of a discipline requires the introduction of
guidelines for emphasizing certain periods, or for smoothing
irregularities in the development. I consider the concepts ``low water
mark'' and the ensuing ``renaissance'' of general relativity as a means
of ordering and valuing events. It was shown here that, unfortunately, these labels are resting on a weak factual basis. Research on general
relativity needed not to be ``resurrected'' or ``reinvigorated'' after
the 1950s: {\em it had yet ``to begin'' at all on a broad scale after world
war II}. The claim of a ``Golden Age'' of general relativity between 1960 and 1975 is a reflection of the growing activity in the field during these years. It is backed by the advent of relativistic astrophysics and a considerable growth of the increase in publications on general relativity during this period.\footnote{The story of a ``Golden Age'' was taken up though by a recent book \cite{Mel2009}. After the erroneously interpreted BICEP-observations, the concept of Golden Age has been applied to hold for our times \cite{Fer2014}; cf. also \cite{Tur2015} in
 another context.} A golden age needs no {\it prior} period of {\em reduced} activity.\\  

The history of general relativity before the 1920s is not taken into
account by many narrations. Besides Einstein and his entourage, and
a few researchers in Berlin, Leiden and Vienna, very few studied general relativity, particularly in France, England and the United States. With the expatriation of Einstein, the focus of research on
general relativity shifted from Germany to other countries, notably to
the United States \cite{Goenner2016}. World War II then slowed down,
or even stopped research for about 25\% of the assumed low-watermark-period. I do not contest that the Bern Jubilee Conference of 1955 gave an impulse to the field of general relativity. It was the {\em first}
international conference on Einstein and his relativity theories since
the introduction of these theories. Now, funds were available for
organizing such conferences: a golden era for ``conference-tourism'' started (cf. appendix 1). The beginnings of relativistic astrophysics, the new impressive variety of communications and technical improvements
leading to the possibility for experimentation, to me seem to be main
factors for establishing general relativity as an accepted subfield of research within the physics community. Yet, the writing of the history of general relativity in the past hundred years needs further detailed studies.\\

\newpage
{\bf Appendix 1}:\\ 
\begin{center}{\bf List of best known national and international conferences in the field of gravitation and cosmology.}\end{center}
\begin{itemize}
\item Jubilee Conference ``50 years of Special Relativity'', Bern 1955.
\item The Conferences GR 1 (1957, Chapel Hill) to GR 21 (2016, New York) 
\item Since 1958 NATO Advanced Study Institutes
\item Enrico Fermi Summer Courses, Varenna. (Gravitation started to be a theme in 1961; then also in 1969, 1972, 1976)
\item ``Symposium on Gravity''  at Research Institute for Fundamental Physics in Japan 1962.
\item Les Houches Summer Schools (University of Grenoble). 1963 ``Relativity, Groups and Topology'' was the 1st about gravity; 
1972 ``Black Holes/Les Astres Occlus'' was the second about gravity.
\item ``Texas Symposium on Relativistic Astrophysics'' from the 1st one in 1963 to the 50th in 2013.
\item Colloques internationaux du Centre National de la Recherche
  Scientifique - No. 167  1967 Fluides et champ gravitationnel en relativit\'e g\'en\'erale; No. 220  1973 Ondes et Radiations Gravitationnelles; No. 236  1974 Th\'eories cin\'etiques classiques et relativistes.
\item Brandeis Summer Institute 1964 (on gravitation). 1968 Astrophysics and Gravitation (11th Brandeis Summer Institute) [Start of the series in 1958]
\item American Mathematical Society Conference 1967 (on gravitation).
\item  International School of Cosmology and Gravitation ``Etore Majorana'',  Erice starting in 1972, another in1979 (NATO advanced study series).
\item Marcel Grossmann-Conferences. From MG 1 in Trieste in 1975 to MG 15 (2015) in Rome. 
\item Edoardo Amaldi Conferences on Gravitational Waves. Begin 1994.
\end{itemize}

\end{document}